\documentclass[aps, prl, twocolumn, floatfix, nofootinbib, superscriptaddress]{revtex4-1}

\usepackage{amsmath,amsfonts,amssymb,bm}
\usepackage{dsfont}
\usepackage{graphicx}
\usepackage{color}
\usepackage{soul}

\usepackage[mathscr,scaled=1.15]{urwchancal}
\DeclareFontFamily{OT1}{pzc}{}
\DeclareFontShape{OT1}{pzc}{m}{it}%
{<-> s * [1.15] pzcmi7t}{}
\DeclareMathAlphabet{\mathpzc}{OT1}{pzc}{m}{it}

\definecolor{purple}{rgb}{0.5,0,0.5}
\definecolor{blue}{rgb}{0.0,0,0.9}
\definecolor{prdblue}{rgb}{0.133,0.118,0.498}
\usepackage[colorlinks=true, pdfstartview=FitV, linkcolor=prdblue, citecolor= prdblue, urlcolor=prdblue]{hyperref}

\begin{document}

\title{New Perspective on Hybrid Mesons}

\author{Shu-Sheng Xu}
\affiliation{Department of Physics, Nanjing University, Nanjing, Jiangsu 210093, China}

\author{Zhu-Fang Cui}
\affiliation{Department of Physics, Nanjing University, Nanjing, Jiangsu 210093, China}

\author{Lei Chang}
\email[]{leichang@nankai.edu.cn}
\affiliation{School of Physics, Nankai University, Tianjin 300071, China}

\author{Joannis Papavassiliou}
\affiliation{Department of Theoretical Physics and IFIC, University of Valencia and CSIC, E-46100, Valencia, Spain}

\author{Craig D. Roberts}
\email[]{cdroberts@anl.gov}
\affiliation{Physics Division, Argonne National Laboratory, Argonne, Illinois
60439, USA}

\author{Hong-Shi Zong}
\email[]{zonghs@nju.edu.cn}
\affiliation{Department of Physics, Nanjing University, Nanjing, Jiangsu 210093, China}
\affiliation{Joint Center for Particle, Nuclear Physics and Cosmology, Nanjing, Jiangsu 210093, China}

\date{23 May 2018}

\begin{abstract}
It is thought that strong interactions within the Standard Model can generate bound-states in which non-Abelian gauge-bosons play a dual role, serving both as force and matter fields.  In this context we introduce a novel approach to the hybrid-meson (valence-gluon+quark+antiquark) bound-state problem in relativistic quantum field theory.  Exploiting the existence of strong two-body correlations in the gluon-quark, $q_g=[gq]$, and gluon-antiquark, $\bar q_g=[g\bar q]$ channels, we argue that a sound description of hybrid properties can be obtained by solving a coupled-pair of effectively two-body equations; and, consequently, that hybrids may be viewed as highly-correlated  $q_g \bar q \leftrightarrow q \bar q_g$ bound-states.  Analogies may be drawn between this picture of hybrid structure and that of baryons, in which quark+quark (diquark) correlations play a key role.  The potential of this formulation is illustrated by calculating the spectrum of light-quark isovector hybrid mesons.  
\end{abstract}



\maketitle

\noindent\emph{1.\,Introduction}\,---\,The known spectrum of light hadrons is simple in the sense that it qualitatively matches the pattern established by constituent-quark models \cite{GellMann:1964nj, Zweig:1981pd}: there are mesons built from a constituent-quark-antiquark ($Q \bar Q$) pair and baryons constituted from three constituent quarks ($QQQ$), where $Q$ is associated with any one of the light $u$-, $d$-, $s$-quarks.  Notably, Refs.\,\cite{GellMann:1964nj, Zweig:1981pd} also raised the possibility that more complicated hadrons are possible, \emph{e.g}.\ $QQ \bar Q\bar Q$ and $Q\bar Q QQQ$.  No candidates were then known; but after fifty years, in systems involving the heavier $c$- and $b$-quarks, that has now changed \cite{Lebed:2016hpi, Ali:2017jda, Olsen:2017bmm}.

Following the ``discovery'' of quantum chromodynamics (QCD) \cite{Marciano:1977su, Marciano:1979wa}, a non-Abelian quantum gauge field theory, in which eight self-interacting gauge bosons (gluons) mediate the interactions between current quarks, new possibilities arose.  Namely, systems with valence glue, \emph{e.g}.\ hybrid mesons, $Q\bar Q G$; hybrid baryons, $QQQG$; and even glueballs, $GG$.  Here, $G$ is a nebulously defined ``constituent gluon'' degree of freedom, whose nature will only become known once such systems are detected experimentally.  Meanwhile, today's tabulations of hadron masses identify at least three plausible hybrid-meson candidates below 2\,GeV \cite{Olive:2016xmw}, and dedicated searches for such states are a worldwide priority, \emph{e.g}.\ Refs.\,\cite{Nerling:2016uwm, Dobbs:2017vjw}.

Numerous models have been employed over time to calculate the spectrum of light hybrid mesons \cite{Lebed:2016hpi, Ali:2017jda, Olsen:2017bmm, Meyer:2015eta, Richard:2016eis}: the approaches are distinguished by, \emph{inter alia}, their disparate treatments/definitions of $G$; and the resulting spectra disagree.  Notwithstanding that, development of a reliable continuum method for calculating hybrid meson properties would be very valuable, primarily for the interpretation of empirical observations but also to provide insights into results obtained via the numerical simulation of lattice-regularised QCD (lQCD) \cite{Dudek:2010wm, Dudek:2011bn}.

$Q\bar Q$ mesons in quantum mechanics cannot possess the following (exotic) quantum numbers \cite{Olive:2016xmw}: $J^{PC} = 0^{+-}$, $0^{--}$, $1^{-+}$, etc.  This is not true in Poincar\'e-covariant treatments of two-valence-body bound states owing to the existence of an additional degree of freedom, \emph{i.e}.\ the relative time between the valence-quark and -antiquark \cite{LlewellynSmith:1969az, Burden:1996nh, Burden:2002ps}.  However, extant studies of exotic mesons using simple, symmetry-preserving \emph{Ans\"atze} or truncations for the Bethe-Salpeter kernel produce unrealistic spectra \cite{Burden:1996nh, Burden:2002ps, Krassnigg:2009zh, Qin:2011dd, Qin:2011xq, Fischer:2014xha, Hilger:2016efh}, \emph{viz}.\ exotic mesons with masses so light that they should already have been seen empirically when, in fact, signals for such states are currently weak and lie at significantly higher masses.  Furthermore, the two-body Bethe-Salpeter equation does not readily entertain a distinction between regular mesons and hybrids with the same $J^{PC}$.  These weaknesses are not remedied when more sophisticated two-body kernels \cite{Chang:2009zb, Chang:2010hb, Chang:2011ei, Williams:2015cvx, Binosi:2016rxz} are used; and we judge them to be a signal that hybrid mesons must contain an explicit valence-gluon degree-of-freedom.

\smallskip

\noindent\emph{2.\,Faddeev Equation for Hybrid Mesons}\,---\,Given the preceding observations, herein we formulate a  Poincar\'e-covariant Faddeev equation for hybrids, which treats these systems as bound-states of a valence-gluon, -quark and -antiquark.  Importantly, each of these valence constituents is strongly dressed, \emph{e.g}.\ possessing an emergent mass that is large at infrared momenta \cite{Roberts:2018hpf}.

Our starting point is an observation \cite{Binosi:2016rxz} that the standard inhomogeneous Bethe-Salpeter equation for the gluon-quark vertex can be rewritten in terms of a gluon-quark (Compton) scattering amplitude, $\mathpzc{C}$, which is one particle irreducible in the $s$-channel:
\begin{subequations}
\label{JPEquation}
\begin{align}
\Gamma_\mu^a(k,q) & = \gamma_\mu t^a  + \tilde \Gamma_\mu^a(k,q) \\
 \tilde \Gamma_\mu^a(k,q) & =
\int^\Lambda_{d\ell}\!\!
\gamma_\rho t^b  S(\ell_+) D_{\rho\sigma}(\bar\ell_-) \,\mathpzc{C}^{ ba}_{\sigma\mu}(\ell,q;k)\,,
\label{qgHBS}
\end{align}
\end{subequations}
where $\{t^a = \lambda^a/2,a=1,\ldots,8\}$, with $\{\lambda^a\}$ the SU$(3)$-colour Gell-Mann matrices; $\int^\Lambda_{d\ell}$ represents a Poincar\'e invariant regularisation of the four-dimensional integral, with $\Lambda$ the regularization mass-scale; and $S$, $D_{\rho\sigma}$ are the dressed-quark and -gluon propagators, respectively, with $\ell_\pm = \ell+\eta k$, $\bar\ell_\pm = \ell \pm (1-\eta) k$, $\eta\in [0,1]$.  Applying charge conjugation, one obtains an equivalent equation for gluon-antiquark scattering.

Recalling textbook derivations of the two-body Bethe-Salpeter equation in analyses of two-particle scattering and the relationship between the scattering matrix and kernel;
and also the role that coloured quark-quark (diquark) correlations play in simplifying the baryon three-body problem \cite{Cahill:1988dx, Burden:1988dt, Cahill:1988zi, Cahill:1987qr, Maris:2002yu, Maris:2004bp, Reinhardt:1989rw, Efimov:1990uz, Eichmann:2009qa},
then Eq.\,\eqref{qgHBS} suggests the possibility that analogous gluon-quark $[gq]$ and degenerate gluon-antiquark $[g\bar q]$ correlations may play a material role in solving the three-body problem for hybrid mesons. 

\begin{figure}[t]
\centerline{%
\includegraphics[clip, width=0.45\textwidth]{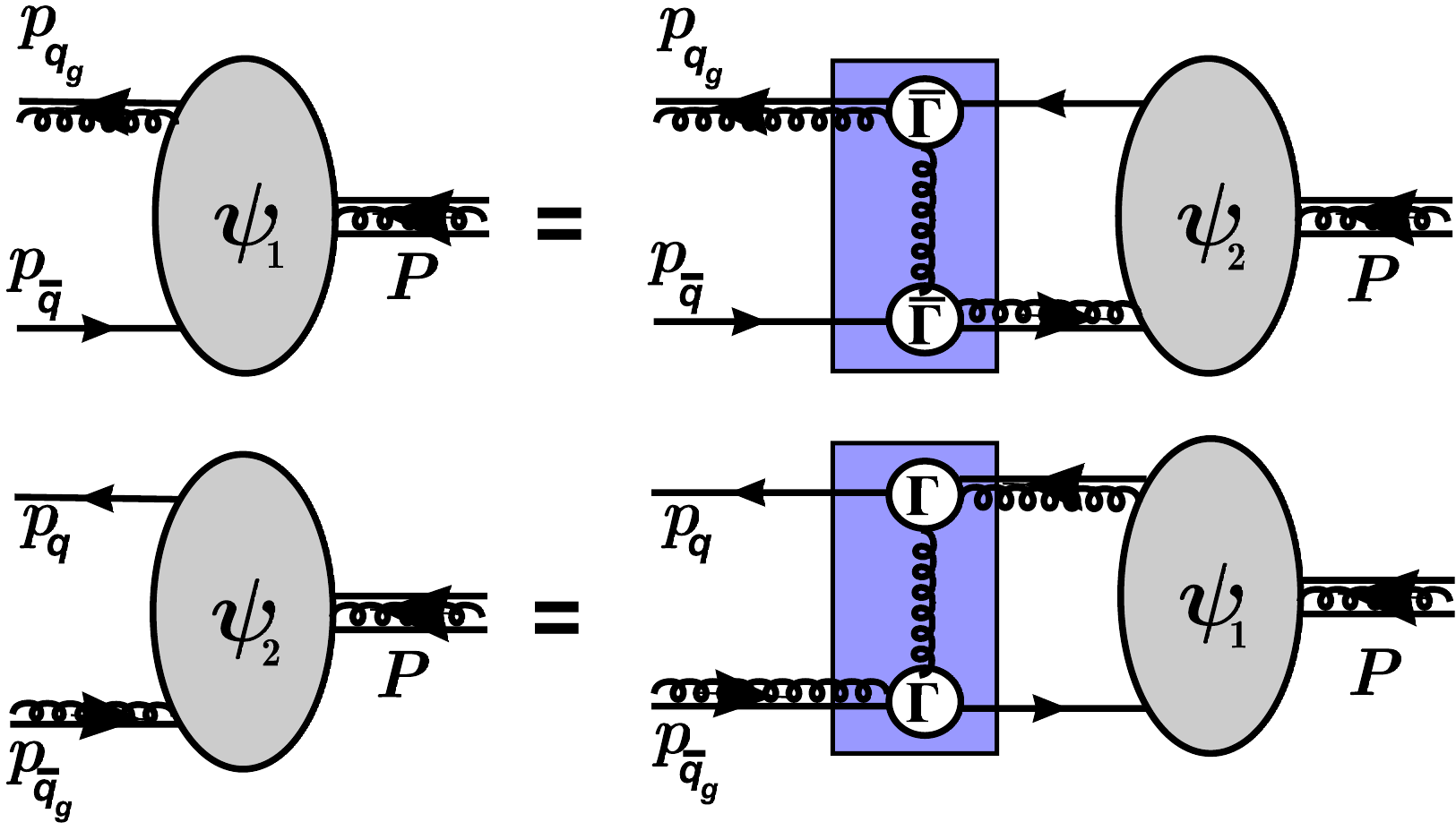}}
\caption{\label{figFaddeev}
Poincar\'e covariant Faddeev equation for a $gq\bar q$ hybrid with total momentum $P=p_{q_g} - p_{\bar q} = p_q + p_g - p_{\bar q} = p_{q} - p_{\bar q_g}$, obtained assuming that strong $gq$ and $g\bar q$ correlations exist within this three-valence-body system.  The complete bound-state amplitude is $\Psi(P) = \Psi_1(P) + \Psi_2(P)$; and the coupled-channels problem simplifies to Eq.\,\eqref{HybridEquation} for states with good charge-con\-ju\-ga\-tion parity.
The rectangles demarcate the active kernels of this Faddeev equation --
single line: dressed quark/antiquark propagator;
adjacent line and spring (coiled line): ${\mathpzc S}_g$, propagators for the $q_g=[gq]$, $\bar q_g=[g\bar q]$ correlations;
$\Gamma$, $\bar\Gamma$: $q_g$, $\bar q_g$ correlation amplitudes, respectively;
and single spring: ${D}$, propagator of the gluon that leaves $\bar q_g$ to combine with $q$, forming $q_g$, and vice versa.}
\end{figure}

Supposing that strong $gq$ and $g\bar q$ correlations exist, then the bound-state equation depicted in Fig.\,\ref{figFaddeev} should provide the basis for a sound description of hybrid mesons.
Here the complete bound-state amplitude is $\Psi = \Psi_1 + \Psi_2$, with $\Psi_{1,2}$ describing the $[q_g\bar q]$, $[q\bar q_g]$ relative momentum correlations, respectively:
\begin{align}
\Psi^a_{1\mu}(p,l;P) = \Gamma^a_{\mu}(l;p_{q_g}) {\mathpzc S}_g(p_{q_g}) \psi_{1}(p;P)\,,
\end{align}
where $\Gamma^a_{\mu}(\ell;p_{q_g})$ is the canonically-normalised amplitude \cite{Nakanishi:1969ph} describing a $[gq]$ correlation with total momentum $p_{q_g}$ and relative momentum $l$, which has the matrix structure of a gluon-quark vertex;
${\mathpzc S}$ is the propagator for this correlation;
and $\psi_{1}(p;P)$ is a meson-like Bethe-Salpeter amplitude.  $\Psi_2$ is plain by analogy.

Focusing on states with good charge-con\-ju\-ga\-tion parity, such as isovector hybrids in the isospin-symmetry limit, then $\psi_{1,2}(p;P) = \bar\psi_{2,1}(p;P)= {\mathpzc p}\psi_{2,1}(p;P) $, ${\mathpzc p} = \pm 1$, where $\bar \psi(p;P) = C [ \psi(-p;P)]^{\rm T} C^\dagger $ with $C=\gamma_2\gamma_4$ the charge-conjugation matrix and $[\cdot]^{\rm T}$ indicating matrix transpose.  Hence, for such systems, the solution of the coupled-channel Faddeev equation in Fig.\,\ref{figFaddeev} is completely determined by the following Bethe-Salpeter equation:
\begin{align}
\nonumber
 \psi_{2}(p;P) & = {\mathpzc p} \! \int_{d\ell} \! D_{\mu\nu}(p_+ - \bar\ell_+)
 \Gamma^a_\mu(p_+ - \tfrac{1}{2}\bar\ell_+;\bar\ell_+){\mathpzc S}_g(\bar \ell_+)\\
& \times \bar \psi_2(\ell;P) S(\ell_-)
 \Gamma^a_\nu(\ell_- - \tfrac{1}{2}\bar p_-;\bar p_-)\,.
 \label{HybridEquation}
\end{align}

The challenges now are to confirm the existence of tight gluon-quark correlations and determine their properties.

\smallskip

\noindent\emph{3.\,Gluon-Quark Correlation}\,---\,
Adapting the logic used to establish the existence and properties of diquark correlations \cite{Cahill:1987qr, Maris:2002yu, Maris:2004bp}, we search for a pole solution to a leading-order (rainbow-ladder \cite{Binosi:2016rxz}) truncation of Eqs.\,\eqref{JPEquation}, \emph{i.e}.\ for a solution of the following homogeneous Bethe-Salpeter equation, $\Gamma_\mu^a = t^a \Gamma_\mu$, $k=p-\ell$:
\begin{align}
\nonumber
t^a & \Gamma_\mu(p;Q) \Lambda_+ = -\int_{d\ell} {\mathpzc G}(k^2)
t^b \gamma_\rho S(\ell_+) \\
& \times  t^c \Gamma_\lambda(\ell;Q) D_{\lambda \tau}(\bar \ell_-) \, {\mathpzc f}_{3g}(k^2)\,
_0\!V_{\rho\tau\mu}^{bca}(k,\bar\ell_- , \bar p_-) \Lambda_+\,.
\label{gqEquation}
\end{align}
where $\Lambda_+ = [-i \gamma\cdot Q + m_{gq}]/[2m_{gq}]$, with $m_{gq}$ the bound-state mass, is the positive-energy projection operator.  Here \cite{Qin:2011dd, Qin:2011xq}
\begin{align}
\label{CalGQC}
\tfrac{1}{Z_2^2}\tilde{\mathpzc G}(k^2) & = \frac{8 \pi^2}{\omega^4} D \, {\rm e}^{-k^2/\omega^2}
+ \frac{8 \pi^2 \gamma_m\, {\mathpzc E}(k^2,m_T^2)}{\ln [ \tau + (1+k^2/\Lambda_{\rm QCD}^2)^2]}\,,
\end{align}
with $\gamma_m = 12/25$, $\Lambda_{\rm QCD}=0.234\,$GeV, $\tau={\rm e}^2-1$, ${\mathpzc E}(k^2,m_T^2) = [1 - \exp(-k^2/m_T^2)]/k^2$, $m_T=2w$, is chosen because it is the basis for a successful description of a wide variety of hadron phenomena; $\! \,_0\!V_{\rho\tau\mu}^{bca}$ is the bare three-gluon vertex; and ${\mathpzc f}_{3g}(k^2) = [1-\exp(-k^2/\omega_{3g}^2)]$, with $\omega_{3g}$ a parameter, is included so that the active three-gluon vertex exhibits the infrared suppression found in modern studies \cite{Aguilar:2013vaa, Athenodorou:2016oyh, Boucaud:2017obn, Corell:2018yil}.  
$Z_2$ is the dressed-quark wave function renormalisation constant: we use a momentum-subtraction scheme, with renormalisation at $\zeta=19\,$GeV \cite{Maris:1997tm}.
(A quark exchange diagram can also contribute to the kernel of Eq.\,\eqref{gqEquation} at this order; but it is numerically sub-leading and therefore neglected.)


In order to solve Eq.\,\eqref{gqEquation}, we first observe that the solution has the form
\begin{subequations}
\begin{align}
\label{Gammagq}
 \Gamma_\mu&(p;Q)  = \sum_{i=1}^6 {\mathpzc g}_i(p;Q) {\mathpzc t}_\mu^i(p,Q)  \,,\\
\nonumber
& {\mathpzc t}^1  = \gamma_\mu\,, \;
{\mathpzc t}^2 =   i \hat p_\mu \,,\;
{\mathpzc t}^3 =  \hat Q_\mu \,, \\
%
& {\mathpzc t}^4 = i \gamma\cdot \hat p  \, \hat Q_\mu \,,
{\mathpzc t}^5 =  i \gamma_\mu \gamma\cdot \hat p\,, \;
{\mathpzc t}^6 =  \gamma\cdot \hat p \,\hat p_\mu \,,
\label{taubasis}
\end{align}
\end{subequations}
with $\hat p ^2 = 1$, $\hat Q^2= -1$, and $\{ {\mathpzc g}_i, i=1,\ldots,6\}$ are scalar functions.  Inserting Eq.\,\eqref{Gammagq} into Eq.\,\eqref{gqEquation} yields a coupled-channels eigenvalue problem for the correlation's  amplitude functions, $\{ {\mathpzc g}_i\}$, and mass-squared, $Q^2=-m_{gq}^2$.  Following Ref.\,\cite{Maris:1997tm}, such equations are now readily solved, and we adapt algorithms from Ref.\,\cite{Krassnigg:2009gd} when necessary.

Completing the kernel of Eq.\,\eqref{gqEquation} with $D\omega = (0.96\,{\rm GeV})^3$, $\omega = 0.5\,$GeV, values determined by fitting an array of $\pi$- and $\rho$-meson observables, with renormalisation group invariant current-quark mass $\hat m = 6.3\,$MeV; $S$ used in calculating those meson results, obtained from the consistent dressed-quark gap equation; $\omega_{3g}=0.9\,$GeV;
and a dressed propagator for the gluon constituent in the form $D_{\lambda \tau}(\bar \ell_-) = \delta_{\lambda \tau}  {\mathpzc E}(\bar \ell_-^2,m_g^2)$, $m_g = 0.6\,$GeV,
one finds a solution of Eq.\,\eqref{Gammagq} with $m_{gq} = 1.0\,$GeV ($=m_{g\bar q}$).  The existence of a solution is robust, \emph{e.g}.\ almost identical results are obtained using the separable model in Refs.\,\cite{Burden:1996nh, Burden:2002ps}.
%

A result $m_{gq} \simeq 1\,$GeV is \emph{natural} because dressed light-quark propagation is characterised by an infrared mass-scale $M \simeq 0.35\,$GeV \cite{Bhagwat:2003vw, Bowman:2005vx, Bhagwat:2006tu}; a dressed-gluon, which couples strongly to a quark-antiquark pair, should have a mass $\simeq 2 M$, which is roughly the value we've chosen and consistent with that computed elsewhere \cite{Binosi:2016xxu}; and the composite $gq$ system should thus have a mass $\simeq 3 M$.  Notably, under combined $\pm 10$\% variations in $\omega_{3g}$ and $m_g$, $m_{gq} \to (1 \pm 0.1) m_{gq}$.

This analysis reveals that, as in colour-antitriplet diquark channels, interactions that deliver a good description of ground-state hadron properties also generate strong $[gq]$ correlations.  They should be confined, but this is not true in RL truncation.  Notably, corrections to this simplest approximation are critical in diquark channels, serving to eliminate bound-state poles from the quark-quark scattering matrix whilst simultaneously preserving the strong diquark correlations \cite{Bhagwat:2004hn}.  Such corrections should similarly affect $[gq]$ correlations.

\smallskip

\noindent\emph{4.\,Hybrid Meson Spectrum}\,---\,Now that Eq.\,\eqref{HybridEquation} is validated, we solve it for the spectrum of  light-quark ground-state hybrids.  Regarding the kernel, having already solved Eq.\,\eqref{gqEquation}, the dressed-quark and -gluon propagators, $S$, $D_{\mu\nu}$, respectively, and canonically-normalised $[gq]$ correlation amplitudes, $\Gamma^a_\nu$, are all in hand.  The remaining element is ${\mathpzc S}_g$.  Eqs.\,\eqref{JPEquation} reveal that a quark-gluon correlation must, in all practical aspects, behave like a dressed quark: it is a colour-triplet fermion-like object whose propagator takes the standard form:
\begin{align}
{\mathpzc S}_g(\ell) = -i \gamma\cdot \ell \sigma_V(\ell^2) + \sigma_S(\ell^2)\,.
\end{align}

One could develop a gap equation for this propagator, which would determine the scalar functions $\sigma_{V,S}$; but that would merely amount to a complicated model.  Herein, we choose instead to be led by experience with dressed quarks and diquark correlations in the baryon problem, and simplify the analysis by employing the following \emph{Ans\"atze} ($s=\ell^2$) \cite{Alkofer:2004yf, Segovia:2014aza}:
\begin{equation}
\label{Eqsigma}
\sigma_V(s)  = {\mathpzc E}(s,{\mathpzc s}_V)  \,,\;
\sigma_S(s)  = \frac{m_{gq}}{s}[ 1 - {\mathpzc s}_S {\mathpzc E}(s,{\mathpzc s}_S) ] \,,
\end{equation}
where the parameters ${\mathpzc s}_{V,S}$ are described in connection with Eq.\,\eqref{params} below.  These forms ensure that ${\mathpzc S}_g$ may be understood to describe a confined excitation owing to the associated violation of reflection positivity \cite{Horn:2016rip}.

As is typical of Faddeev equations involving composite degrees of freedom, the $[gq]$ correlation amplitudes in Eq.\,\eqref{HybridEquation} are sampled off-shell and the integrand therefore exhibits spurious ultraviolet behaviour unless one introduces an off-shell extension of the basis in Eq.\,\eqref{taubasis}.  Adapting the procedure in Ref.\,\cite{Eichmann:2007nn}, we write $\hat Q = Q h(Q^2) /[i m_{gq}]$, ${\mathpzc t}^{i\neq 1} \to {\mathpzc t}^{i\neq 1} h(Q^2)^2$, where $h(Q^2)^2 = m_{gq}^2/[Q^2+2 m_{gq}^2]$.  Evidently, $h(Q^2=-m_{gq}^2) = 1$.
%

\begin{table}[t]
\caption{\label{TableA}
Row~1: Hybrid-meson masses obtained from Eq.\,\eqref{HybridEquation}.  Under combined $\pm$10\% changes of $m_{qg}$ and ${\mathpzc s}_{V,S}$ in Eq.\,\eqref{Eqsigma}, these results have the indicated sensitivity.
Row~2: Hybrid spectrum obtained after enhancing the anomalous chromomagnetic moment term in the $[gq]$ correlation amplitude, see Eq.\,\eqref{params}: $\kappa_{gq} \to \kappa_{gq} (1 \pm 0.1)$ has negligible impact.
The lQCD results in Rows~5 and 6 were computed on dynamical lattices with pion masses $m_\pi \approx (0.52, 0.44, 0.40)\,$GeV and two volumes $16^3\times 128$, $20^3\times 128$  \cite{Dudek:2010wm}.  These configurations yield a mass for the pion's first radial excitation that is approximately $\delta_{\pi_1} = 0.43$\,GeV larger than experiment \cite{Olive:2016xmw}.  In Rows~3 and 4, we renormalise the lQCD results by subtracting $\delta_{\pi_1}$.
(All tabulated results listed in GeV.)}
\begin{center}
\begin{tabular*}
{\hsize}
{
l@{\extracolsep{0ptplus1fil}}
|c@{\extracolsep{0ptplus1fil}}
c@{\extracolsep{0ptplus1fil}}
c@{\extracolsep{0ptplus1fil}}
c@{\extracolsep{0ptplus1fil}}
c@{\extracolsep{0ptplus1fil}}}
                         & $0^{-+}$ & $1^{-+}$ & $1^{--}$ & $0^{+-}$ & $0^{--}$ \\ \hline
RL direct             & 1.28(9)    & 1.80(4)     & 1.64(10)     & 1.73(13)     & 1.74(3) \\
ACM improved\;  & 1.62(6)    & 1.75(8)     & 1.86(10)     & 1.87(14)     & 1.90(3) \\\hline
lQCD$_R$ - $16^3$    & 1.72(2)    & 1.73(2)     & 1.84(2)$\phantom{0}$      & 2.03(1)$\phantom{0}$      & \\
lQCD$_R$ - $20^3$    & 1.69(2)    & 1.72(2)     & 1.77(6)$\phantom{0}$      & 1.99(2)$\phantom{0}$  &  \\
lQCD - $16^3$    & 2.14(1)    & 2.15(2)     & 2.26(2)$\phantom{0}$      & 2.45(1)$\phantom{0}$      & \\
lQCD - $20^3$    & 2.12(2)    & 2.16(2)     & 2.21(6)$\phantom{0}$      & 2.43(2)$\phantom{0}$  &  \\
\hline
\end{tabular*}
\end{center}
\end{table}

The kernel of Eq.\,\eqref{HybridEquation} is now complete; and using the approach just described and the full bound-state amplitude in all cases \cite{LlewellynSmith:1969az, Krassnigg:2009zh}, we solved Eq.\,\eqref{HybridEquation} for the ground-state masses and on-shell amplitudes in the following five light-quark isovector channels: $J^{PC} = 0^{-+}$, $1^{--}$, $0^{+-}$, $0^{--}$, $1^{-+}$, the last three of which are quark model exotics.  %
(We used $\eta=1/3$ because it simplifies the numerical analysis and the masses do not depend on this choice when Poincar\'e-covariance is preserved throughout.)
The spectrum thus obtained is listed in Row~1 of Table~\ref{TableA}.  Bound-states exist in all channels with, notably, the $0^{-+}$, $1^{--}$ hybrids being structurally distinct from those accessible using the two-body Bethe-Salpeter equation in the channels.
However, in comparison with lQCD predictions \cite{Dudek:2010wm}, almost all states are too light, especially $0^{-+}$, and the $1^{-+}$-$1^{--}$ ordering is reversed.  Wide variations of $m_{gq}$, $\mathpzc{s}_{V,S}$, do not alter this outcome.

The mismatch between the spectrum in Row~1 of Table~\ref{TableA} and lQCD results caused us to reconsider each element in our formulation of the hybrid meson problem.  Drawing upon analyses of improvements to RL truncation \cite{Chang:2009zb, Chang:2010hb, Chang:2011ei}, we were led to the probable origin.  Namely, the $[gq]$ correlation amplitude was computed in RL truncation and, consequently, the anomalous chromomagnetic moment (ACM) associated with this correlation is greatly underestimated owing to the failure of RL truncation to reliably express effects of dynamical chiral symmetry breaking \cite{Chang:2010hb} (DCSB or, equivalently, emergence of strong mass).  We therefore omitted the spin-independent coupling ${\mathpzc t}^3$, multiplied the ACM related term in Eq.\,\eqref{taubasis}, ${\mathpzc t}^5$, by a parameter, $\kappa_{gq}$, and varied ($\kappa_{gq}$, $\mathpzc{s}_{V,S}$) in search of a triplet that reproduces the sign of the $1^{-+}$-$\,1^{--}$-$\,0^{+-}$ lQCD splittings.  Notably, no constraint was placed on the value of any single mass.  The fact that a solution to this problem does exist:
\begin{equation}
\label{params}
\kappa_{gq} = 2.4\,,\;
{\mathpzc s}_V = (0.70 \, {\rm GeV})^2\,,\;
{\mathpzc s}_S = (0.63\,{\rm GeV})^2\,,
\end{equation}
supports our conjecture regarding the origin of the mismatch; and these values generate the spectrum in Row~2 of Table~\ref{TableA}.  (This analysis suggests that no solution of the valence-gluon+quark+anti\-quark Faddeev equation which adheres strictly to the leading-order RL truncation can produce a realistic spectrum of hybrids.)

\begin{figure}[t]
\centerline{%
\includegraphics[clip, width=0.45\textwidth]{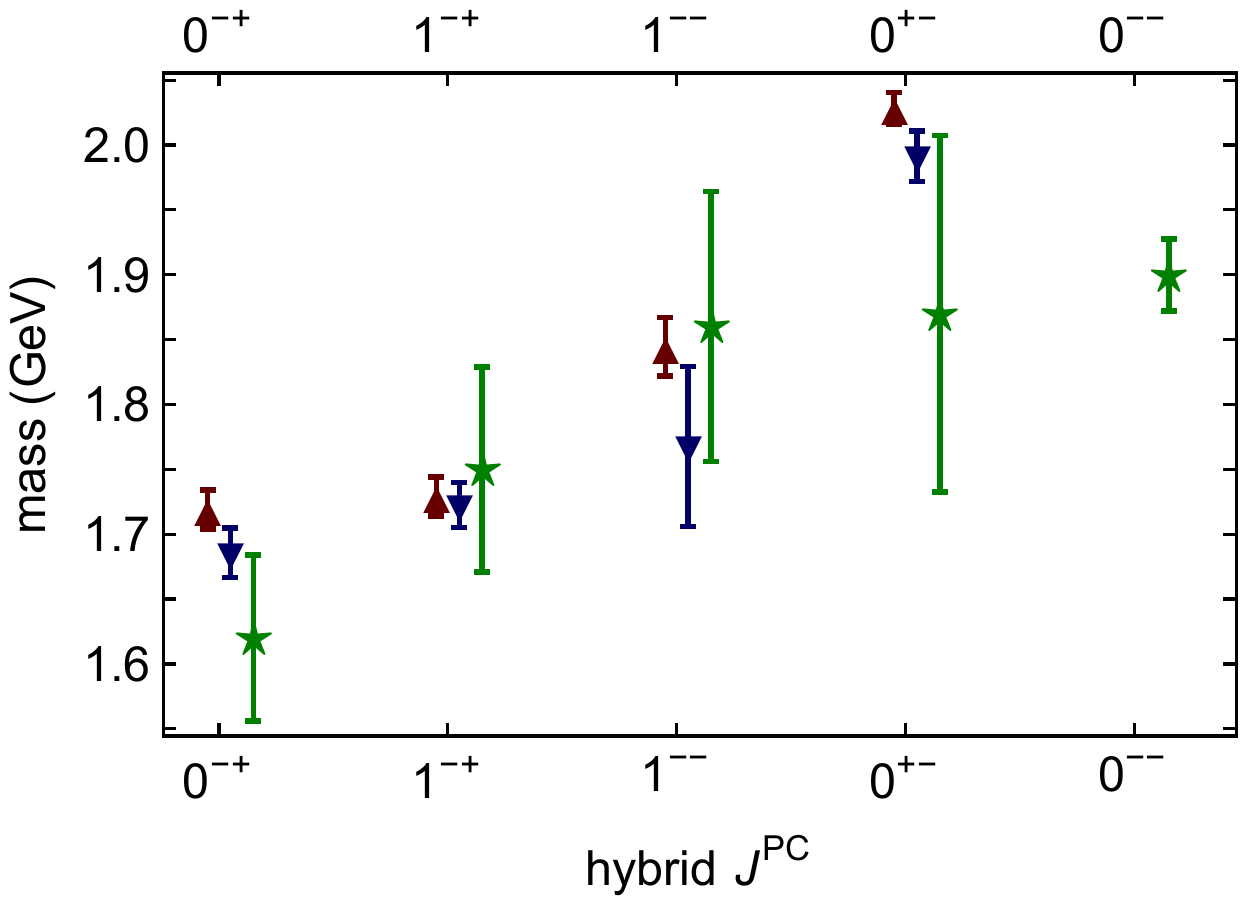}}
\caption{\label{spectrumplot}
Comparison between our ACM-improved spectrum (stars, green), Row~2 in Table~\ref{TableA}, and the rescaled lQCD results in Rows~3 (up-triangles, red) and 4 (down-triangles, blue).}
\end{figure}

Figure~\ref{spectrumplot} compares the lattice results in Rows~3, 4 of Table~\ref{TableA} with our ACM-improved calculations in Row~2.  The level ordering is identical.  Furthermore, the absolute values of the masses are commensurate.  This is a nontrivial outcome.  The lQCD masses plotted in Fig.\,\ref{spectrumplot} are rescaled by subtraction of $\delta_{\pi_1}$, the difference between the simulation results for the mass of the pion's first radial excitation and its empirical value, a number which is completely unrelated to our calculations.  Moreover, as remarked above, no single lQCD mass was used as a constraint when fitting the parameters in Eq.\,\eqref{params}.  The magnitude of our results is instead set by the infrared values of the running gluon and quark masses, which are determined by $\pi$- and $\rho$-meson properties, unrelated to hybrid channels.

The $0^{--}$ state deserves additional attention.  lQCD predicts this channel to host the heaviest light-quark hybrid ground-state, with the lightest such meson lying more than 2\,GeV above the ground-state $\rho$-meson \cite{Dudek:2010wm}.   The Faddeev equation in Fig.\,\ref{figFaddeev}, which capitalises on the existence and properties of $[gq]$, $[g\bar q]$ correlations in $gq\bar q$ scattering, confirms that the $0^{--}$ is the heaviest hybrid, thereby correcting a defect of RL-truncation analyses of exotics using the two-body Bethe-Salpeter equation \cite{Qin:2011dd, Qin:2011xq, Fischer:2014xha, Hilger:2016efh}.  Our computed $0^{--}$ mass might nevertheless be too light because such a system probably possesses a large amount of angular momentum \cite{Berwein:2015vca}, leading to significant DCSB-enhanced repulsion within the bound-state; and our simple expedient for ameliorating the associated defects of RL truncation may not be completely adequate.  The approach we have described will always produce a heavy $0^{--}$ state, but its precise location must await future, more sophisticated analyses.

\smallskip

\noindent\emph{5.\,Epilogue}\,---\,%
We introduced a novel approach to the valence-gluon+quark+antiquark bound-state problem in relativistic quantum field theory.  Beginning with the associated Faddeev equation, we demonstrated that strong correlations exist in the gluon-quark, $q_g=[gq]$, and gluon-antiquark, $\bar q_g=[g\bar q]$ channels; and hence that a simpler, coupled pair of effectively two-body equations can provide the basis for a realistic description of hybrid mesons [Fig.\,\ref{figFaddeev}].  Consequently, hybrid mesons appear as highly-correlated  $q_g \bar q \leftrightarrow q \bar q_g$ bound-states.  In hindsight, given the role that coloured quark+quark correlations play in determining the properties of the neutron, proton, and other baryons, the existence and importance of kindred correlations within hybrids appears credible.

To illustrate the potential of this formulation, we employed a simple model to describe the properties of the $[gq]$, $[g\bar q]$ correlations and therewith established that it can readily reproduce the location and ordering of ground-state light-quark hybrids obtained via the numerical simulation of lattice-regularised QCD.  This outcome distinguishes our approach from other continuum frameworks.  Notably, an understanding of the impact of emergent strong-interaction mass in forming the $[gq]$, $[g\bar q]$ correlations is crucial to this success.

Ours is a first analysis of hybrid mesons from this new perspective.  More sophisticated treatments are necessary before the validity of the formulation can firmly be established.  Meanwhile, it should serve as a guide for subsequent continuum treatments of the hybrid-meson three-body problem; and the highly-correlated wave functions it yields can be used to predict a range of hybrid decays and other processes in order to elucidate empirical signatures for the presence and role of $[gq]$, $[g\bar q]$ correlations.

\smallskip

%
\noindent\emph{Acknowledgments}\,---\,
%
We are grateful for constructive suggestions from D.~Binosi, S.-X.~Qin and J.~Segovia.
Work supported by:
China Postdoctoral Science Foundation (under Grant No.\ 2016M591809);
the Chinese Government's \emph{Thousand Talents Plan for Young Professionals};
Spanish MEyC, under grants FPA2017-84543-P and SEV-2014-0398;
Generalitat Valenciana  under grant Prometeo~II/2014/066;
the Chinese Ministry of Education, under the \emph{International Distinguished Professor} programme;
U.S.\ Department of Energy, Office of Science, Office of Nuclear Physics, under contract no.~DE-AC02-06CH11357;
and National Natural Science Foundation of China (under Grant Nos.\ 11475085, 11535005 and 11690030).


\begin{thebibliography}{10}

\bibitem{GellMann:1964nj}
M.~Gell-Mann,
\newblock Phys. Lett. {\bf 8}, 214 (1964).

\bibitem{Zweig:1981pd}
G.~Zweig,
\newblock (1964),
\newblock {\emph{An $SU(3)$ model for strong interaction symmetry and its
  breaking. Parts 1 and 2} (CERN Reports No.\ 8182/TH.\ 401 and No.\ 8419/TH.\
  412)}.

\bibitem{Lebed:2016hpi}
R.~F. Lebed, R.~E. Mitchell and E.~S. Swanson,
\newblock Prog. Part. Nucl. Phys. {\bf 93}, 143 (2017).

\bibitem{Ali:2017jda}
A.~Ali, J.~S. Lange and S.~Stone,
\newblock Prog. Part. Nucl. Phys. {\bf 97}, 123 (2017).

\bibitem{Olsen:2017bmm}
S.~L. Olsen, T.~Skwarnicki and D.~Zieminska,
\newblock Rev. Mod. Phys. {\bf 90}, 015003 (2018).

\bibitem{Marciano:1977su}
W.~J. Marciano and H.~Pagels,
\newblock Phys. Rept. {\bf 36}, 137 (1978).

\bibitem{Marciano:1979wa}
W.~J. Marciano and H.~Pagels,
\newblock Nature {\bf 279}, 479 (1979).

\bibitem{Olive:2016xmw}
C.~Patrignani {\em et~al.},
\newblock Chin. Phys. C {\bf 40}, 100001 (2016).

\bibitem{Nerling:2016uwm}
F.~Nerling,
\newblock EPJ Web Conf. {\bf 126}, 04033 (2016),
\newblock {\emph{Highlights from the COMPASS experiment at CERN}}.

\bibitem{Dobbs:2017vjw}
S.~Dobbs,
\newblock PoS {\bf Hadron2017}, 047 (2018),
\newblock {\emph{Searching for Hybrid Mesons with GlueX}}.

\bibitem{Meyer:2015eta}
C.~A. Meyer and E.~S. Swanson,
\newblock Prog. Part. Nucl. Phys. {\bf 82}, 21 (2015).

\bibitem{Richard:2016eis}
J.-M. Richard,
\newblock Few Body Syst. {\bf 57}, 1185 (2016).

\bibitem{Dudek:2010wm}
J.~J. Dudek, R.~G. Edwards, M.~J. Peardon, D.~G. Richards and C.~E. Thomas,
\newblock Phys. Rev. D {\bf 82}, 034508 (2010).

\bibitem{Dudek:2011bn}
J.~J. Dudek,
\newblock Phys. Rev. D {\bf 84}, 074023 (2011).

\bibitem{LlewellynSmith:1969az}
C.~H. Llewellyn-Smith,
\newblock Annals Phys. {\bf 53}, 521 (1969).

\bibitem{Burden:1996nh}
C.~J. Burden, L.~Qian, C.~D. Roberts, P.~C. Tandy and M.~J. Thomson,
\newblock Phys. Rev. C {\bf 55}, 2649 (1997).

\bibitem{Burden:2002ps}
C.~J. Burden and M.~A. Pichowsky,
\newblock Few Body Syst. {\bf 32}, 119 (2002).

\bibitem{Krassnigg:2009zh}
A.~Krassnigg,
\newblock Phys. Rev. D {\bf 80}, 114010 (2009).

\bibitem{Qin:2011dd}
S.-X. Qin, L.~Chang, Y.-X. Liu, C.~D. Roberts and D.~J. Wilson,
\newblock Phys. Rev. C {\bf 84}, 042202(R) (2011).

\bibitem{Qin:2011xq}
S.-X. Qin, L.~Chang, Y.-X. Liu, C.~D. Roberts and D.~J. Wilson,
\newblock Phys. Rev. C {\bf 85}, 035202 (2012).

\bibitem{Fischer:2014xha}
C.~S. Fischer, S.~Kubrak and R.~Williams,
\newblock Eur. Phys. J. A {\bf 50}, 126 (2014).

\bibitem{Hilger:2016efh}
T.~Hilger and A.~Krassnigg,
\newblock Eur. Phys. J. A {\bf 53}, 142 (2017).

\bibitem{Chang:2009zb}
L.~Chang and C.~D. Roberts,
\newblock Phys. Rev. Lett. {\bf 103}, 081601 (2009).

\bibitem{Chang:2010hb}
L.~Chang, Y.-X. Liu and C.~D. Roberts,
\newblock Phys. Rev. Lett. {\bf 106}, 072001 (2011).

\bibitem{Chang:2011ei}
L.~Chang and C.~D. Roberts,
\newblock Phys. Rev. C {\bf 85}, 052201(R) (2012).

\bibitem{Williams:2015cvx}
R.~Williams, C.~S. Fischer and W.~Heupel,
\newblock Phys. Rev. D {\bf 93}, 034026 (2016).

\bibitem{Binosi:2016rxz}
D.~Binosi, L.~Chang, S.-X. Qin, J.~Papavassiliou and C.~D. Roberts,
\newblock Phys. Rev. D {\bf 93}, 096010 (2016).

\bibitem{Roberts:2018hpf}
C.~D. Roberts,
\newblock (2018),
\newblock {\emph{N* Structure and Strong QCD}, arXiv:1801.08562 [nucl-th]}.

\bibitem{Cahill:1988dx}
R.~T. Cahill, C.~D. Roberts and J.~Praschifka,
\newblock Austral. J. Phys. {\bf 42}, 129 (1989).

\bibitem{Burden:1988dt}
C.~J. Burden, R.~T. Cahill and J.~Praschifka,
\newblock Austral. J. Phys. {\bf 42}, 147 (1989).

\bibitem{Cahill:1988zi}
R.~T. Cahill,
\newblock Austral. J. Phys. {\bf 42}, 171 (1989).

\bibitem{Cahill:1987qr}
R.~T. Cahill, C.~D. Roberts and J.~Praschifka,
\newblock Phys. Rev. D {\bf 36}, 2804 (1987).

\bibitem{Maris:2002yu}
P.~Maris,
\newblock Few Body Syst. {\bf 32}, 41 (2002).

\bibitem{Maris:2004bp}
P.~Maris,
\newblock Few Body Syst. {\bf 35}, 117 (2004).

\bibitem{Reinhardt:1989rw}
H.~Reinhardt,
\newblock Phys. Lett. B {\bf 244}, 316 (1990).

\bibitem{Efimov:1990uz}
G.~V. Efimov, M.~A. Ivanov and V.~E. Lyubovitskij,
\newblock Z. Phys. C {\bf 47}, 583 (1990).

\bibitem{Eichmann:2009qa}
G.~Eichmann, R.~Alkofer, A.~Krassnigg and D.~Nicmorus,
\newblock Phys. Rev. Lett. {\bf 104}, 201601 (2010).

\bibitem{Nakanishi:1969ph}
N.~Nakanishi,
\newblock Prog. Theor. Phys. Suppl. {\bf 43}, 1 (1969).

\bibitem{Aguilar:2013vaa}
A.~C. Aguilar, D.~Binosi, D.~Iba{\~n}ez and J.~Papavassiliou,
\newblock Phys. Rev. D {\bf 89}, 085008 (2014).

\bibitem{Athenodorou:2016oyh}
A.~Athenodorou {\em et~al.},
\newblock Phys. Lett. B {\bf 761}, 444 (2016).

\bibitem{Boucaud:2017obn}
P.~Boucaud, F.~De~Soto, J.~Rodr{\'i}guez-Quintero and S.~Zafeiropoulos,
\newblock Phys. Rev. D {\bf 95}, 114503 (2017).

\bibitem{Corell:2018yil}
L.~Corell, A.~K. Cyrol, M.~Mitter, J.~M. Pawlowski and N.~Strodthoff,
\newblock (arXiv:1803.10092 [hep-ph]),
\newblock {\emph{Correlation functions of three-dimensional Yang-Mills theory
  from the FRG}}.

\bibitem{Maris:1997tm}
P.~Maris and C.~D. Roberts,
\newblock Phys. Rev. C {\bf 56}, 3369 (1997).

\bibitem{Krassnigg:2009gd}
A.~Krassnigg,
\newblock PoS {\bf CONFINEMENT8}, 075 (2008).

\bibitem{Bhagwat:2003vw}
M.~S. Bhagwat, M.~A. Pichowsky, C.~D. Roberts and P.~C. Tandy,
\newblock Phys. Rev. C {\bf 68}, 015203 (2003).

\bibitem{Bowman:2005vx}
P.~O. Bowman {\em et~al.},
\newblock Phys. Rev. D {\bf 71}, 054507 (2005).

\bibitem{Bhagwat:2006tu}
M.~S. Bhagwat and P.~C. Tandy,
\newblock AIP Conf. Proc. {\bf 842}, 225 (2006).

\bibitem{Binosi:2016xxu}
D.~Binosi, C.~D. Roberts and J.~Rodr{\'i}guez-Quintero,
\newblock Phys. Rev. D {\bf 95}, 114009 (2017).

\bibitem{Bhagwat:2004hn}
M.~S. Bhagwat, A.~H{\"o}ll, A.~Krassnigg, C.~D. Roberts and P.~C. Tandy,
\newblock Phys. Rev. C {\bf 70}, 035205 (2004).

\bibitem{Alkofer:2004yf}
R.~Alkofer, A.~H{\"o}ll, M.~Kloker, A.~Krassnigg and C.~D. Roberts,
\newblock Few Body Syst. {\bf 37}, 1 (2005).

\bibitem{Segovia:2014aza}
J.~Segovia, I.~C. Clo{\"e}t, C.~D. Roberts and S.~M. Schmidt,
\newblock Few Body Syst. {\bf 55}, 1185 (2014).

\bibitem{Horn:2016rip}
T.~Horn and C.~D. Roberts,
\newblock J. Phys. G. {\bf 43}, 073001 (2016).

\bibitem{Eichmann:2007nn}
G.~Eichmann, A.~Krassnigg, M.~Schwinzerl and R.~Alkofer,
\newblock Annals Phys. {\bf 323}, 2505 (2008).

\bibitem{Berwein:2015vca}
M.~Berwein, N.~Brambilla, J.~Tarrús~Castellà and A.~Vairo,
\newblock Phys. Rev. D {\bf 92}, 114019 (2015).

\end{thebibliography}

\end{document}